\documentclass[fleqn,usenatbib]{mnras}

\usepackage{newtxtext,newtxmath}
\usepackage[T1]{fontenc}
\usepackage{ae,aecompl}

\usepackage{graphicx}	% Including figure files
\usepackage{amsmath}	% Advanced maths commands
\usepackage{amssymb}	% Extra maths symbols
\usepackage{xspace}
\usepackage{color}
\usepackage{units}
\usepackage{soul} % strikethough with \st

\title[New mass bound on fermionic dark matter]{New mass bound on fermionic dark matter from a combined analysis of classical dSphs}

\author[D. Savchenko \& A. Rudakovskyi]{
D. Savchenko$^{1}$\thanks{E-mail: dsavchenko@bitp.kiev.ua}
and 
A. Rudakovskyi$^{1}$
\\
$^{1}$Bogolyubov Institute of Theoretical Physics, Metrolohichna Str. 14-b, 03143, Kyiv, Ukraine
}

\date{Accepted XXX. Received YYY; in original form ZZZ}

\pubyear{2019}

\begin{document}
\label{firstpage}
\pagerange{\pageref{firstpage}--\pageref{lastpage}}
\maketitle

\begin{abstract}
Dwarf spheroidal galaxies (dSphs) are the most compact dark matter-dominated objects observed so far. The Pauli exclusion principle limits the number of fermionic dark matter particles that can compose a dSph halo. This results in a well-known lower bound on their particle mass. So far, such bounds were obtained from the analysis of individual dSphs. In this paper, we model dark matter halo density profiles via the semi-analytical approach 
and analyse the data from eight `classical' dSphs assuming the same mass of dark matter fermion in each object. First, we find out that modelling of Carina dSph results in a much worse fitting quality compared to the other seven objects. From the combined analysis of the kinematic data of the remaining seven `classical' dSphs, we obtain a new $2\sigma$ lower bound of $m\gtrsim 190$\,eV on the dark matter fermion mass. In addition, by combining a sub-sample of four dSphs -- Draco, Fornax, Leo~I and Sculptor -- we conclude that 220\,eV fermionic dark matter appears to be preferred over the standard CDM at about 2$\sigma$ level. However, this result becomes insignificant if all seven objects are included in the analysis. Future improvement of the obtained bound requires more detailed data, both from `classical' and ultra-faint dSphs. 
\end{abstract}

\begin{keywords}
dark matter -- galaxies: haloes -- galaxies: dwarf -- galaxies: kinematics and dynamics -- methods: statistical
\end{keywords}

%%%%%%%%%%%%%%%%%%%%%%%%%%%%%%%%%%%%%%%%%%%%%%%%%%

%%%%%%%%%%%%%%%%% BODY OF PAPER %%%%%%%%%%%%%%%%%%
\section{Introduction}
The nature of dark matter (DM) is one of the major questions in modern physics. The mass of DM particle candidates, which exist in numerous extensions of the Standard Model, varies in very wide range -- from $\sim10^{-22}$\,eV for ultra-light DM  \citep[e.g.,][]{Hu:00,Hui:2016ltb,Lee:18} up to TeVs for WIMPs \citep[see, e.g.,][and references therein]{Roszkowski:2017nbc,Arcadi:2017kky}  or up to $\sim10^{13}$\,GeV for WIMPZILLAs \citep[e.g.][]{Chung:1998zb}.  

The Pauli principle forbids packing too many fermions into a gravitationally bound object. Therefore, the average phase-space density of such an object with mass $M$ enclosed within a region of radius $R$,  $\bar{F}\sim \frac{M}{R^3\sigma^3}$, cannot exceed some maximum $f_\text{max}(m)$, where $m$ is the mass of fermion, and $\sigma$ is the particle velocity dispersion. This allows one to obtain the lower bound $m\simeq0.5$\,keV  \citep{Bode:00,Dalcanton:00,Boyarsky:08a,Horiuchi:13},
based on the extended Tremaine--Gunn \citep{Tremaine:79} approach \citep[see also][]{Gorbunov:08,Shao:12,Wang:2017} from the analysis of compact DM dominated objects -- dwarf spheroidal satellites (dSphs). This approach requires an estimator of the dynamical mass $M$ within a sphere of some radius $R$~\citep{Wolf:09,Walker:11,Campbell:16}, see also \citet{Kowalczyk:12} for a detailed study of the mass estimator uncertainties, and \citet{Boyarsky:08a} for the estimate of the phase-space volume occupied by the DM particles.  

Another method for constraining the mass of the fermionic DM particle uses direct comparison between the detailed prediction of the kinematics of dSph and the observational data \citep[see, e.g.,][]{Domcke:2014kla,Randall:16,DiPaolo:2017geq}. It does not require an estimate of the averaged phase-space density over a spatial region. Direct modelling of kinematics also allows one to incorporate the anisotropy of the velocity dispersion into analysis. Moreover, unlike the Tremaine--Gunn approach, this method allows one to combine the data on several objects to produce better limits on the particle mass. 
In return, it requires a (semi-)analytical model of the DM density profile and stellar density profile. Many analytical models of fermionic DM halo density profiles have been developed so far; see, e.g., \citet{Ruffini:83,Bilic:97,Angus:09,deVega:13,deVega:14,Merafina:14, Domcke:2014kla,Ruffini:14,Chavanis:14b,Arguelles:16,Randall:16, Rudakovskyi:18rs, Giraud:2018gxl,Barranco:2018gjg}. 

In this paper, we present a new lower bound on the mass of fermionic DM particle, based on the observed kinematics  \citep{Bonnivard:2015xpq} and photometry \citep{McConnachie:12} data of `classical' dSphs, and assuming the DM density model of~\citet{Rudakovskyi:18rs}. 
In comparison to  \citet{Domcke:2014kla}, \citet{DiPaolo:2017geq}, this approach allows us not only to analyse individual dSphs, but also to perform \emph{combined} statistical analysis based on the total $\chi^2$ goodness-of-fit statistics assuming the same dark matter particle mass in all of them. Thereby, we aim at utilising fully the statistical power of the approach.

This paper is organised as follows: in Sec.~\ref{sec:methods} our methods are described (a short description of our model of fermionic DM halo is also included), the obtained results are summarised in Sec.~\ref{sec:results} and discussed in Sec.~\ref{sec:discussion}. We use the recent \emph{Planck} \citep{Planck-2018} cosmological parameters for our calculations.

\section{Methods}\label{sec:methods}
We use the semi-analytical method proposed in~\cite{Rudakovskyi:18rs} to obtain the density profile of a dark matter halo. It predicts a cored halo for the general case of warm fermionic dark matter without any extra assumptions about the particle model. 
Here we briefly summarise this method. 

For a fermionic dark matter model with particle mass $m_\mathrm{DM}$ and $g$ initial degrees of freedom (hereafter, we assume $g = 2$) the phase space density cannot exceed~\citep{Boyarsky:08a} 
\begin{equation}
    f_{\text{max}} = \frac{g m_\text{DM}^4}{2(2\pi\hbar)^3}.
\end{equation} 
For a steady-state isotropic spherically symmetrical dark matter halo (see~\cite{Rudakovskyi:18rs} for a discussion on the applicability of this assumption) the phase space density $f$ is obtained by using the Eddington transformation~\citep{Eddington:1916,Binney-Tremaine:08book} 
\begin{equation}
    f(\mathcal{E}) = \frac{1}{\pi^2\sqrt{8}}\frac{d}{d\mathcal{E}}\int_{\mathcal{E}}^{0} \frac{d\rho}{d\Phi}\frac{d\Phi}{\sqrt{\mathcal{E}-\Phi}}\,,
\end{equation}
where $\Phi$ is the local gravitational potential. We perform the iterative procedure starting from the NFW profile and truncating the phase space density so that it does not exceed the limiting value: 
\begin{equation}
    f_\text{tNFW}(\mathcal{E})\, {=}\, \left \{ 
    \begin{array}{ll}
     \!\!f(\mathcal{E}) , &   f(\mathcal{E}) < f_\text{max} \, ,\\
         \!\! f_\text{max}, & f(\mathcal{E}) \geq f_\text{max} \, .
        \end{array} \right. 
\end{equation}
After this, we reconstruct the mass density~\citep{Binney-Tremaine:08book}
\begin{equation}
    \rho_\text{tNFW}(r) = 4\pi\int_{\Phi(r)}^0 f_\text{tNFW}(\mathcal{E})\sqrt{2\left(\mathcal{E} - \Phi(r)\right)} d\mathcal{E}
\end{equation} 
for the subsequent step. \cite{Rudakovskyi:18rs} shows good convergence of this procedure after several iterations. We call the obtained profile tNFW (stands for truncated Navarro--Frenk--White) hereafter. The density profiles obtained in this model are in a good agreement with numerical $N$-body simulations \citep{Shao:12,Maccio:12b,Maccio:12err}, see more in \cite{Rudakovskyi:18rs}.
 
Given the density distribution of a dark matter halo, we follow the logic of \cite{Domcke:2014kla} and \cite{DiPaolo:2017geq} to obtain the velocity dispersion along the line of sight. Specifically, we solve the spherical Jeans equation for the radial velocity dispersion $\sigma_\mathrm{r}$, 
 \begin{equation}\label{eq:sigmar}
     \left(\frac{\partial}{\partial r} + \frac{2\beta}{r}\right) (n_\star \sigma_\mathrm{r}^2) = -n_\star \frac{G M(r)}{r^2}\, ,
 \end{equation}
 with the stellar velocity dispersion anisotropy $\beta=1-\sigma^2_\perp/\sigma^2_\mathrm{r}$. In the above, $M(r)$ is the dark matter mass distribution, and $n_\star$ is the stellar number density, which we represent by the  Plummer profile~\citep{Plummer:1911}
 \begin{equation}
     n_\star(r)=n_0\left(1+r^2/r_\mathrm{h}^2\right)^{-5/2}.
 \end{equation}
 The half-light radii $r_h$ for the objects of interest were taken from  \cite{McConnachie:12} and are given in Table~\ref{tab:objdata}.
 We then calculate the velocity dispersion along the line of sight:
\begin{equation}\label{eq:sigmalos}
    \sigma_\mathrm{los}^2(R)=\frac{1}{\Sigma_\star}\int_{R^2}^{\infty}\text{d}r^2\frac{n_\star}{\sqrt{r^2-R^2}}\sigma_\mathrm{r}^2\left[1-\beta \frac{R^2}{r^2}\right],
\end{equation}
where $\Sigma_\star(R)=\int_{R^2}^{\infty}\text{d}r^2n_\star(r)/\sqrt{r^2-R^2}$ \citep{Binney-Tremaine:08book, DiPaolo:2017geq}.
 
We model the binned data on the velocity dispersion for eight classical dSphs taken from~\cite{Bonnivard:2015xpq}. For every mass of the dark matter particle in the 100\,eV -- 900\,eV range with logarithmic split we use brute-force grid optimisation over the tNFW profile parameters $c_{200}$, $M_{200}$, and velocity dispersion anisotropy $\beta$ to minimise the objective $\chi^2$ statistics
\begin{equation}
    \chi^2 = \sum_{i} \frac{\left(\sigma_\mathrm{los,obs}(r_i)-\sigma_\mathrm{los,th}(r_i)\right)^2}{\delta^2(r_i)}\, ,
\end{equation}
where $\sigma_\mathrm{los,obs}(r_i)$ denotes the $i$'th observational point, $\delta^2(r_i)$ is its $1\sigma$ error, and the $\sigma_\mathrm{los,th}(r_i)$ is the predicted value at this point; the summation is performed over the observational points.

\section{Results}\label{sec:results}
\begin{figure*}
    \centering
    \includegraphics[width=0.97\textwidth]{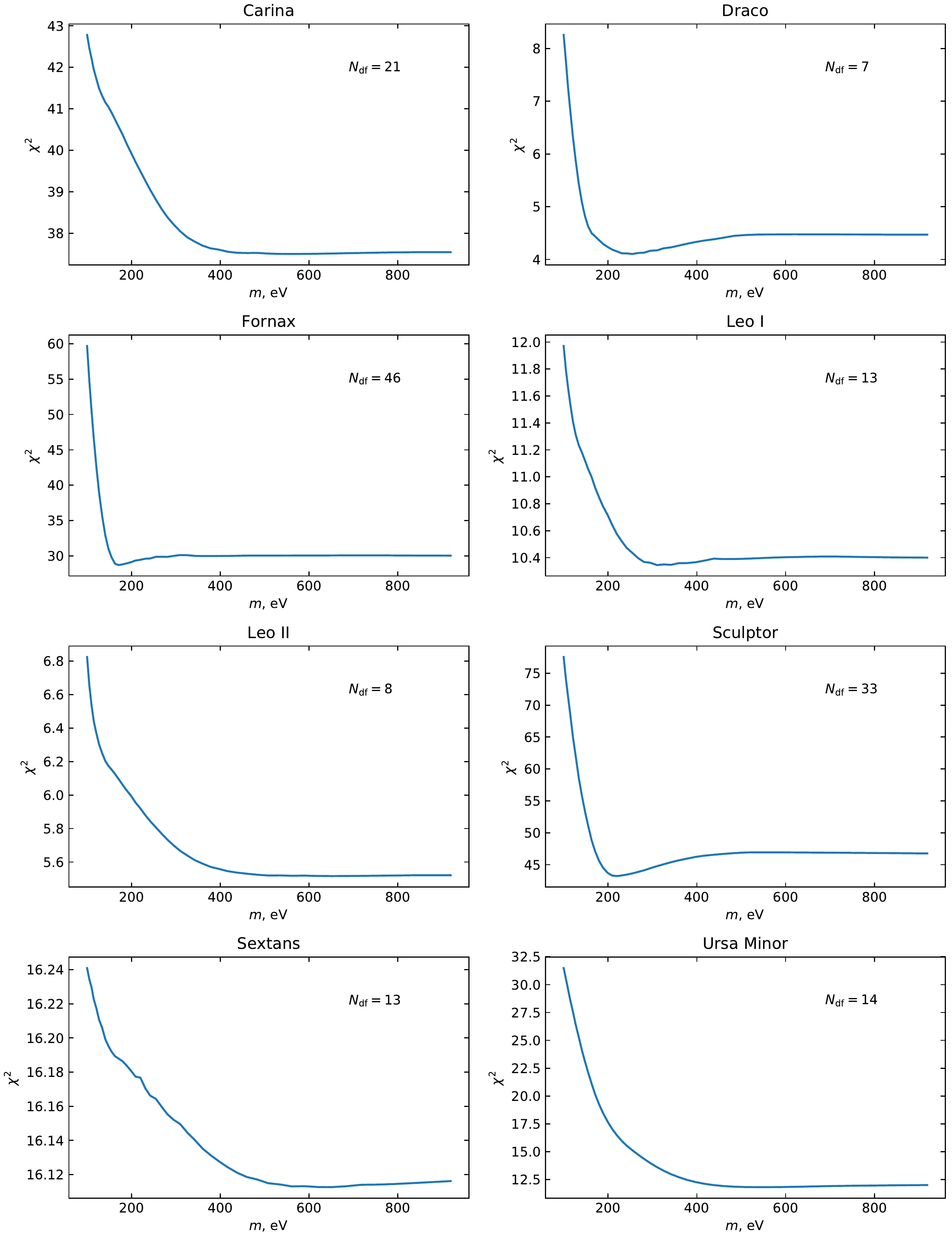}
    \caption{Minimal values of $\chi^2$ statistics as functions of the DM particle mass for the tNFW profile model for each of the eight `classical' dSphs studied in this paper. Also given is the number of the degrees of freedom of the fits. Notice the clearly visible minimums in four objects: Draco, Fornax, Leo I, Sculptor.}
    \label{fig:tnfw_byobj}
\end{figure*}

The dependence of the best-fitting $\chi^2$ statistics on the particle mass for every individual object is plotted in Fig.~\ref{fig:tnfw_byobj}, and the best-fitting model parameters are summarised in Table~\ref{tab:objdata}.

\begin{table*}
     \centering
     \begin{tabular}{cccccccc}
          Object & $r_\mathrm{h}$, kpc & $\chi^2/N_\mathrm{df}$ & $m_\text{DM}$, eV & $M_{200},\,10^{8}M_\odot$ & $c_{200}$ & $\beta$ & $r_\mathrm{c}$, kpc\\
          \hline
          Carina & 0.25 & 37.5/21 & 561 & 111.7 & 5 & 0.21 & 0.25\\
          Draco  & 0.221 & 4.1/7 & 255 & 177.8 & 10 & 0.34 & 0.66\\
          Fornax & 0.71 & 28.7/46 & 171 & 9.57 & 53 & -0.05 & 0.93\\
          Leo1 & 0.251 & 10.4/13 & 310 & 155.7 & 8 & 0.44 & 0.54\\
          Leo2 & 0.176 & 5.5/8 & 650 & 127.6 & 9 & 0.61	& 0.17\\
          Sculptor & 0.283 & 43.2/33 & 220 & 6.01 &	59 & 0.10 & 0.59\\
          Sextans & 0.695 & 16.1/13 & 650 & 875.6 &	2 &	-0.38 & 0.22\\
          Ursa Minor & 0.181 & 11.8/14 & 561 & 4.92 & 36 & -1.32 & 0.15\\
     \end{tabular}
     \caption{The best-fitting parameter values for the modelled objects and goodness-of-fit statistics. Profile parameters correspond to the Navarro-Frenk-White profile used as the starting one in the tNFW generation procedure. Also provided is the half-light radii used in our fits. Core radii $r_c$ for density profiles with the best-fitting parameters, are calculated from $\rho_\text{tNFW}(r_\mathrm{c})=\frac{\rho_\text{tNFW}(0)}{4}$
     according to the definition in~\citet{Rudakovskyi:18rs}.}
     \label{tab:objdata}
 \end{table*}

The goodness-of-fit is acceptable for every object except Carina dSph, which is the only dSph from our selection that has best-fitting $\chi^2$ higher than two standard deviations ($2\sqrt{2N_\mathrm{df}}$) above the mean value $\chi^2_\mathrm{mean}=N_\mathrm{df}$ of the chi-squared distribution. Therefore, we exclude Carina dSph from the subsequent combined analysis. 

Apart from the individual fits, we are interested in the combined goodness-of-fit. We consider the overall $\chi^2$ to be the sum of chi-squared statistics of the individual fits for every dark matter particle mass. The overall best fit is obtained for the particle mass of 342\,\unit{eV} with $\chi^2=124.7$ for 134 degrees of freedom. This value of mass, however, cannot be statistically distinguished from the higher values, as the differences between the corresponding chi-squares are negligible. For comparison, we fitted the data using the Navarro--Frenk--White profile~\citep{Navarro:95,Navarro:96}, typical to the standard CDM dark matter model. The best-fitting $\chi^2$ statistics is 125.1 for 134 degrees of freedom, so none of this models is preferred by our analysis.

\begin{figure}
    \centering
    \includegraphics[width=\columnwidth]{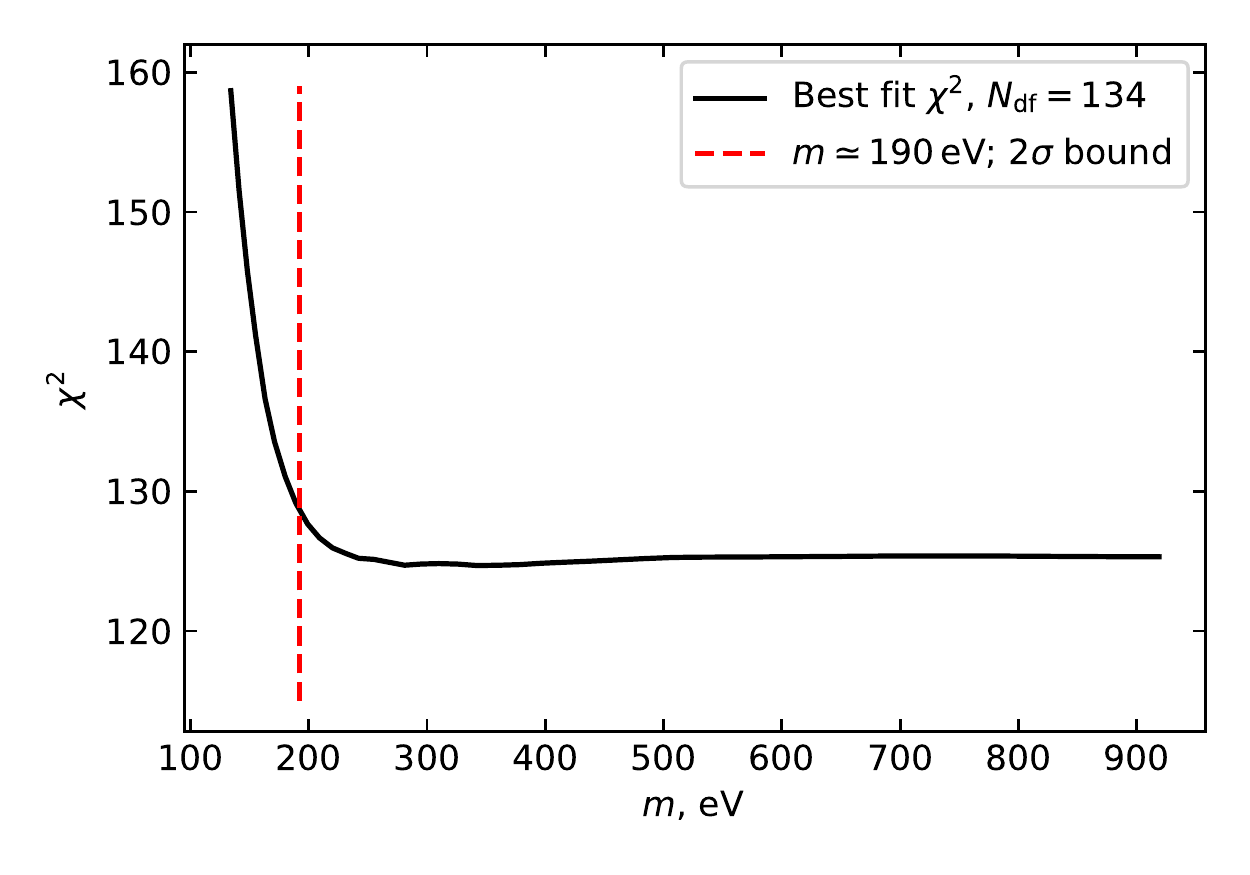}
    \caption{Overall best-fitting $\chi^2$ statistics as a function of the dark matter particle mass. In the limit of high mass the curve approaches the value obtained in the fit with the Navarro--Frenk--White profile, as the tNFW halo model approaches that of the NFW in this limit. 
    %Horizontal blue line is where the level of chi-squared statistics equals the number of dimensions of freedom (mean value of the chi-squared distribution). 
    The dashed line shows the $2\sigma$ confidence bound on the particle mass.} 
    \label{fig:tnfw_chi2}
\end{figure}

\begin{figure}
    \centering
    \includegraphics[width=\columnwidth]{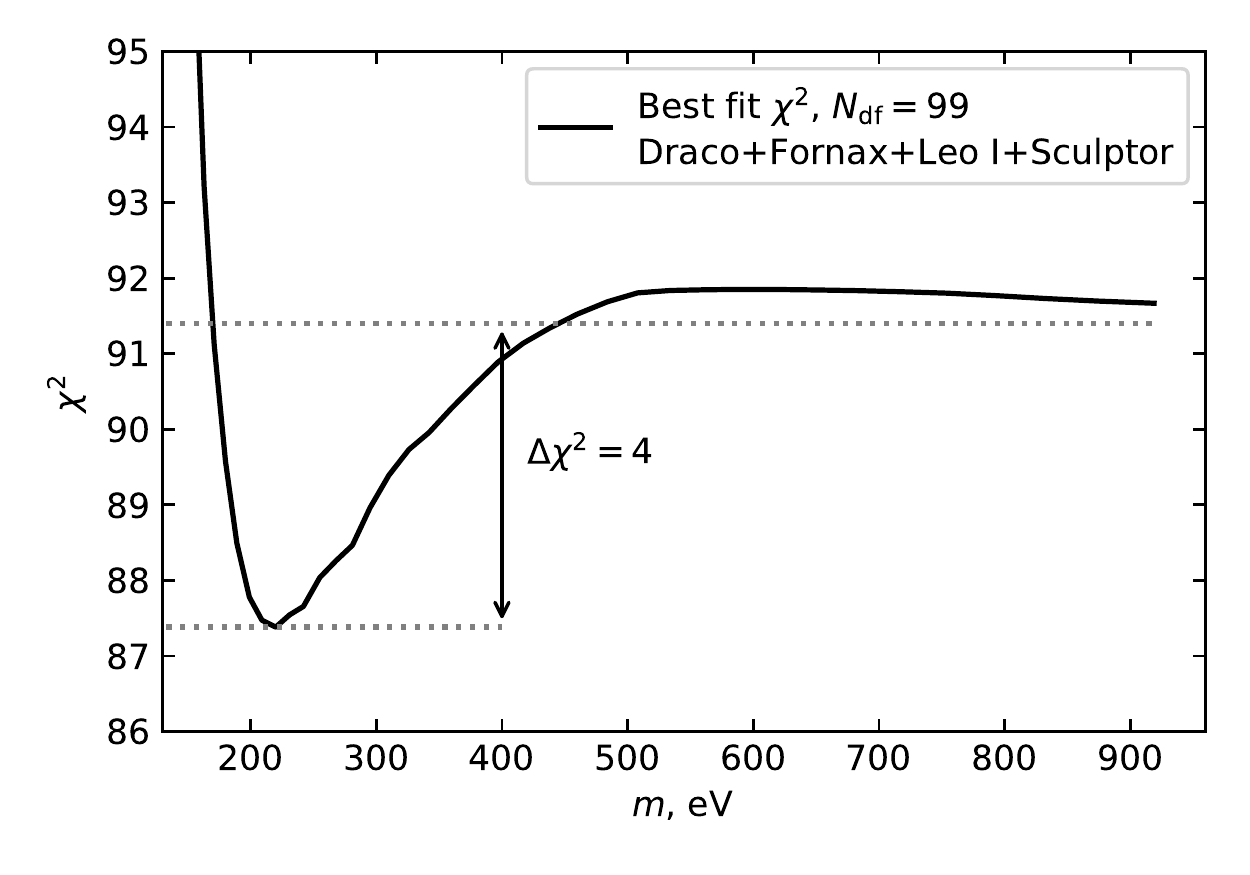}
    \caption{Overall best-fitting $\chi^2$ as a function of dark matter particle mass for the combined analysis of only four selected objects: Draco, Fornax, Leo1 and Sculptor. The minimum at 220\,eV indicating the preferred particle mass is clearly visible. The depth of the dip corresponds to $2\sigma$ significance ($\Delta\chi^2=4$). However, it becomes negligibly small ($\Delta\chi^2=0.4$) when the rest of objects are included into analysis, see Fig.~\protect\ref{fig:tnfw_chi2}.} 
    \label{fig:tnfw_4obj}
\end{figure}

Using the dependence of the overall best-fitting statistics on the particle mass, we can build the confidence range for the mass via the standard approach, described in Sec.~15.6 of~\citet{Press:07book}. The lower bound on the particle mass is the value for which $\chi^2=\chi^2_\text{best-fit}+\Delta\chi^2$, where for $2\sigma$ confidence level $\Delta\chi^2 = 4$. The resulting mass bound of $m_{2\sigma}\simeq190$\,eV is shown in Fig.~\ref{fig:tnfw_chi2}.  

\begin{figure*}
    \centering
    \includegraphics[width=0.89\textwidth]{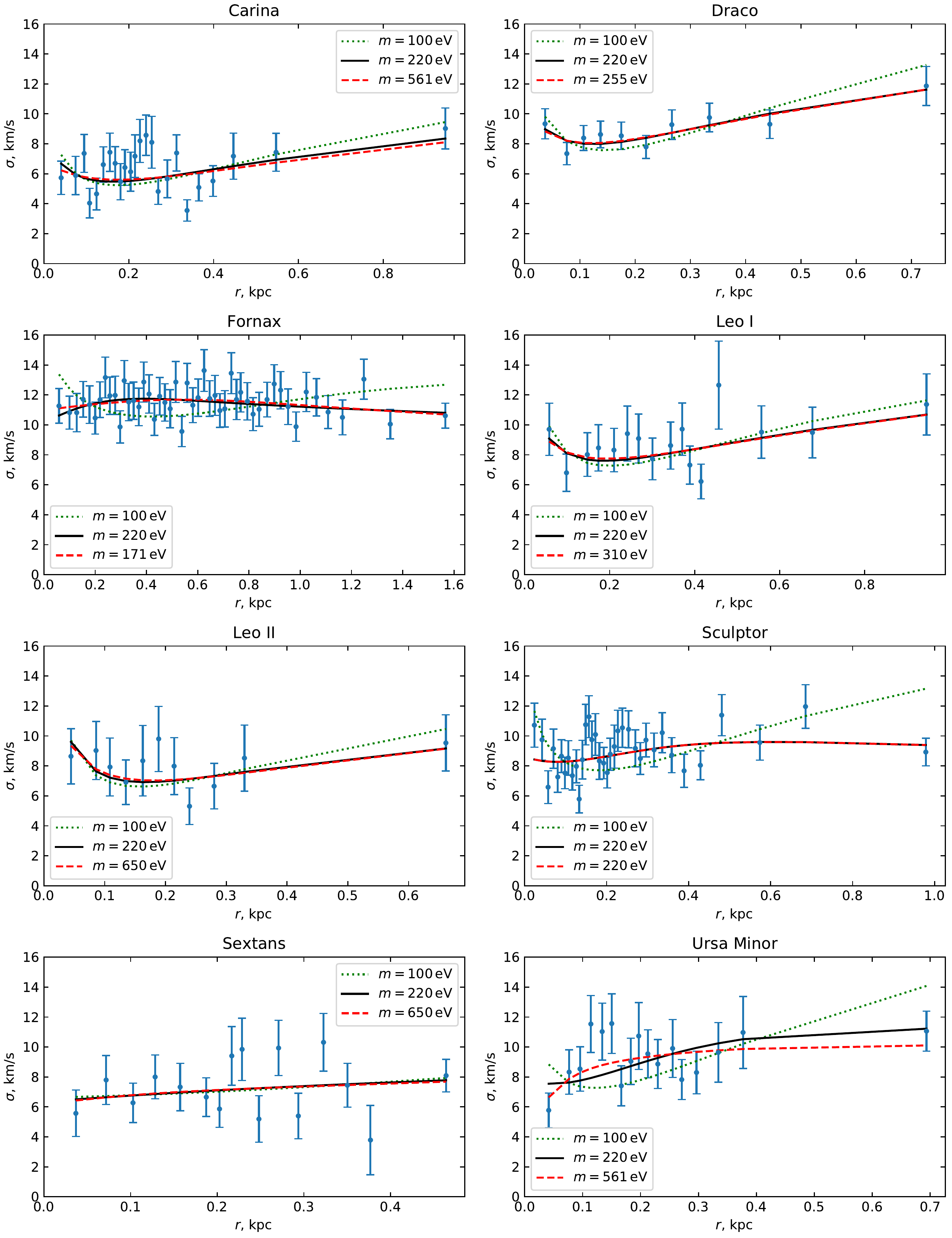}
    \caption{Velocity dispersion along the line of sight versus the distance from the object centre. The dots with error bars represent the binned observational data, taken from~\protect\citet{Bonnivard:2015xpq}. The lines show the best-fitting dependence obtained in the tNFW model with different particle masses, namely, $m=100$\,eV which is below the obtained $2\sigma$ bound, $m=$220\,eV preferred by the combined analysis of four selected dSphs (Draco, Fornax, Leo I, Sculptor), and the mass which provides the minimal $\chi2$ in the fit of the corresponding individual object. One can see that for most of objects the behaviour of $\sigma_\mathrm{los}$ in the case of low particle mass strongly changes whereas variation of mass above the 190\,eV bound has small impact on this behaviour.}
    \label{fig:tnfw_sigma}
\end{figure*}

In Fig.~\ref{fig:tnfw_sigma} we show the effect of particle mass on the velocity dispersion profile in all objects. It is clearly seen that small particle masses strongly modify this profile.

We also combine four objects that show notable local minimum on $\chi^2$ vs mass dependence, namely, Draco, Fornax, Leo1 and Sculptor. The combined fitting statistics in this case is plotted in Fig.~\ref{fig:tnfw_4obj}. The minimum $\chi^2_\mathrm{min}=87.4$ for 99 degrees of freedom is obtained for particle mass $m=220$\,eV, whereas the Navarro--Frenk--White profile fits the data with $\chi^2=91.4$. Thus one can conclude that 220\,eV fermionic dark matter is preferred over CDM with $\Delta\chi^2=4$. However, this observation should not be treated as a strict result, because inclusion of the rest of objects into analysis reduces the local minimum to a statistically insignificant depth of $\Delta\chi^2=0.4$.

\section{Conclusions \& Discussion}\label{sec:discussion}

In this paper, we derive a new maximally model-independent bound on the mass of fermionic dark matter particle. We use the halo model of \citet{Rudakovskyi:18rs} and the Jeans equation for modelling the line-of-sight velocity dispersion. We obtain the conservative $2\sigma$ lower bound $m\gtrsim 190$\,eV on the mass of fermionic dark matter particle. Fermionic DM with higher particle mass cannot be distinguished from the CDM. This result is based on the analysis of seven `classical' dSphs. Our model fails to fit the kinematics of the Carina dSph. However, this galaxy shows the strongest signs of a tidal disruption among the other `classical' dSphs \citep{Munoz:06, Munoz:08, Battaglia:12, Battaglia:2013wqa, Fabrizio:16}, see also \citet{McMonigal:14}. It appears that Carina was transformed from a disky galaxy to a spheroidal via strong tidal interaction with Milky Way \citep{Fabrizio:16}, and different sub-populations have  different kinematic patterns \citep{Fabrizio:16, Hayashi:2018uop}.

To check the robustness of the obtained result with respect to possible uncertainties in the values of half-light radii, we repeat the analysis using upper and lower confidence bounds on $r_\text{h}$ reported by~\citet{McConnachie:12}. We found that the obtained lower bound on the particle mass changes by less than 10\%, being lower for the higher $r_h$ used in the model and vice-versa.

Using only the data on four selected objects, namely, Draco, Fornax, Leo1 and Sculptor, we obtain that fermionic DM with $m=220$\,eV particle mass is preferred over CDM on $2\sigma$ level. The significance decreases to a negligible value when the rest of objects are included into the analysis.

Conceptually, the halo built from the low-mass fermions has an extended core with low central density compared to the cases of more massive DM particles. The best-fitting halos in case of fermions with the mass $m=100$\,eV show $\sim2$\,\unit{kpc} cores for all the objects in the analysis. This is much larger than the radial spans of the outermost points of the observable kinematics.

\begin{figure*}
    \centering
    \includegraphics[width=\textwidth]{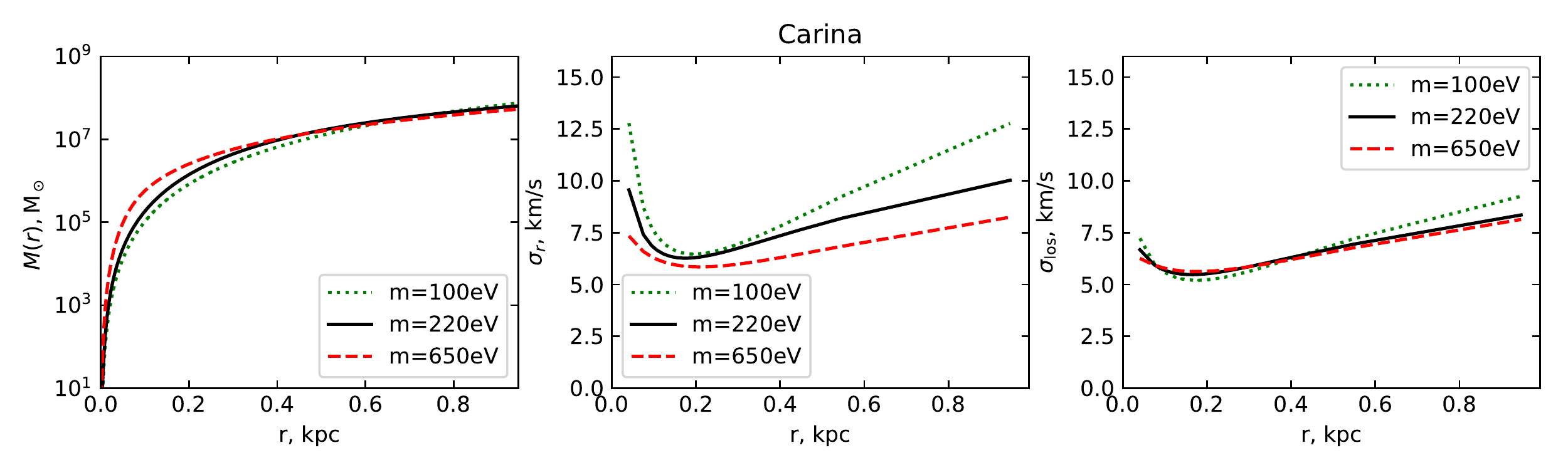}
    \includegraphics[width=\textwidth]{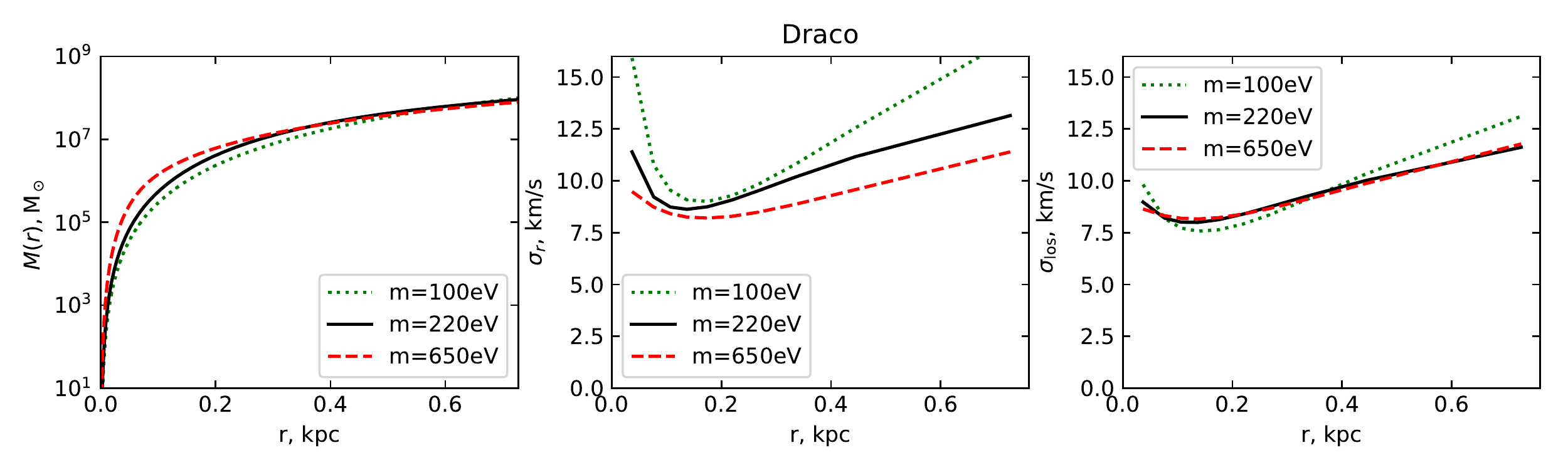}
    \includegraphics[width=\textwidth]{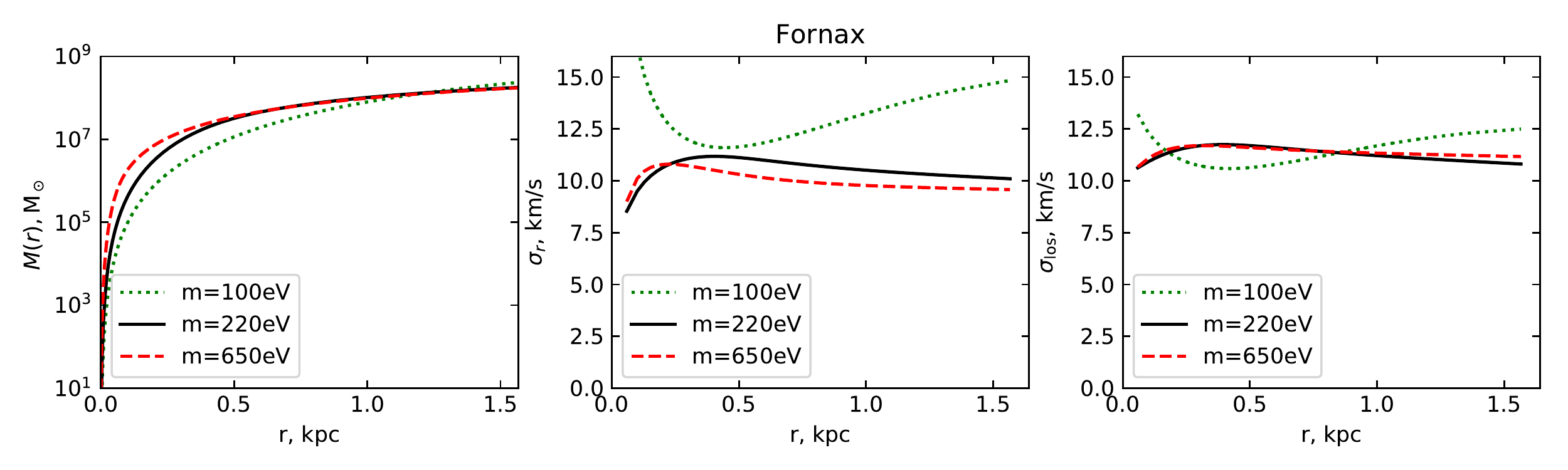}
    \includegraphics[width=\textwidth]{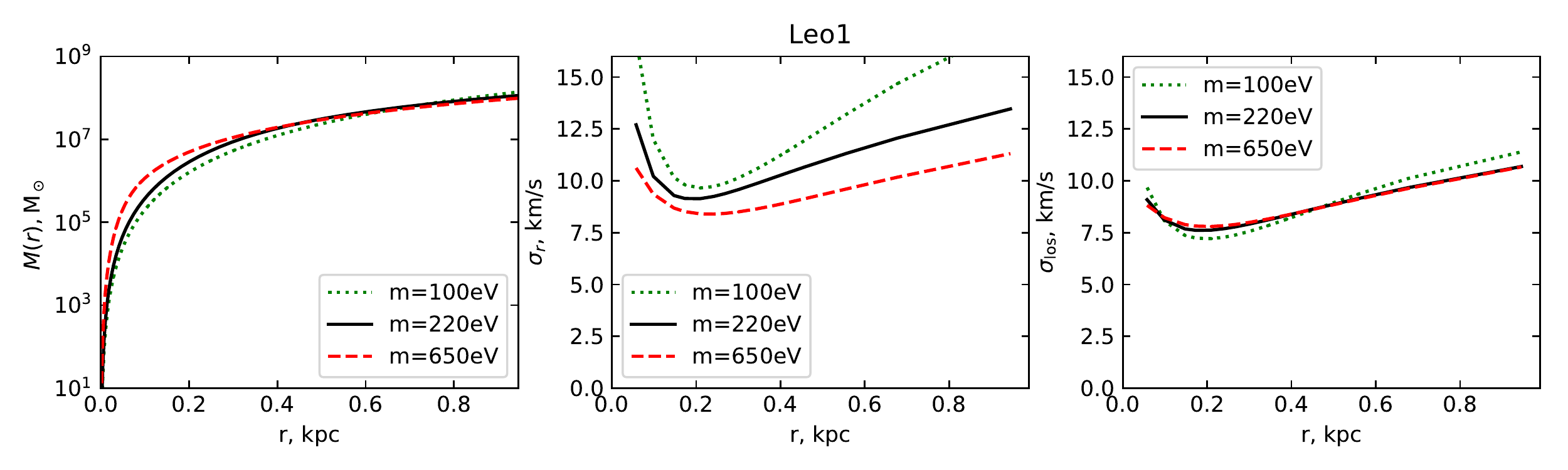}
    \caption{The dependence of the enclosed mass, radial velocity anisotropy, line-of-sight velocity anisotropy from the off-center distance for the three fixed particle masses, same as in Table~\ref{tab:corerh}. First four dSphs from the analysis shown.} 
    \label{fig:mass-mr-mlos-1}
\end{figure*}

\begin{figure*}
    \centering
    \includegraphics[width=\textwidth]{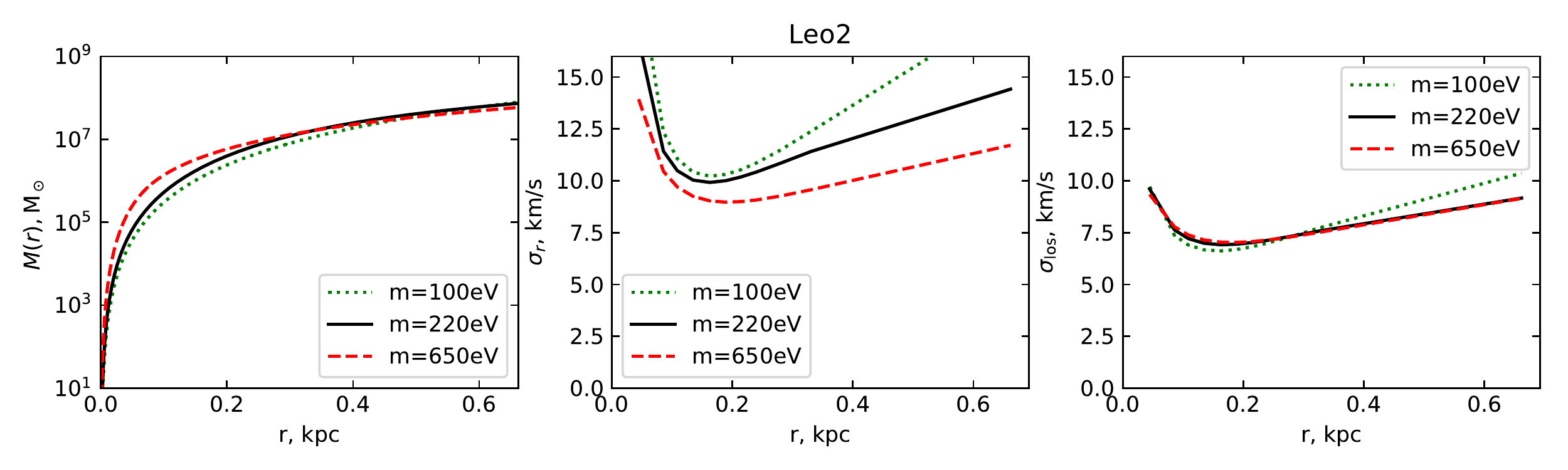}
    \includegraphics[width=\textwidth]{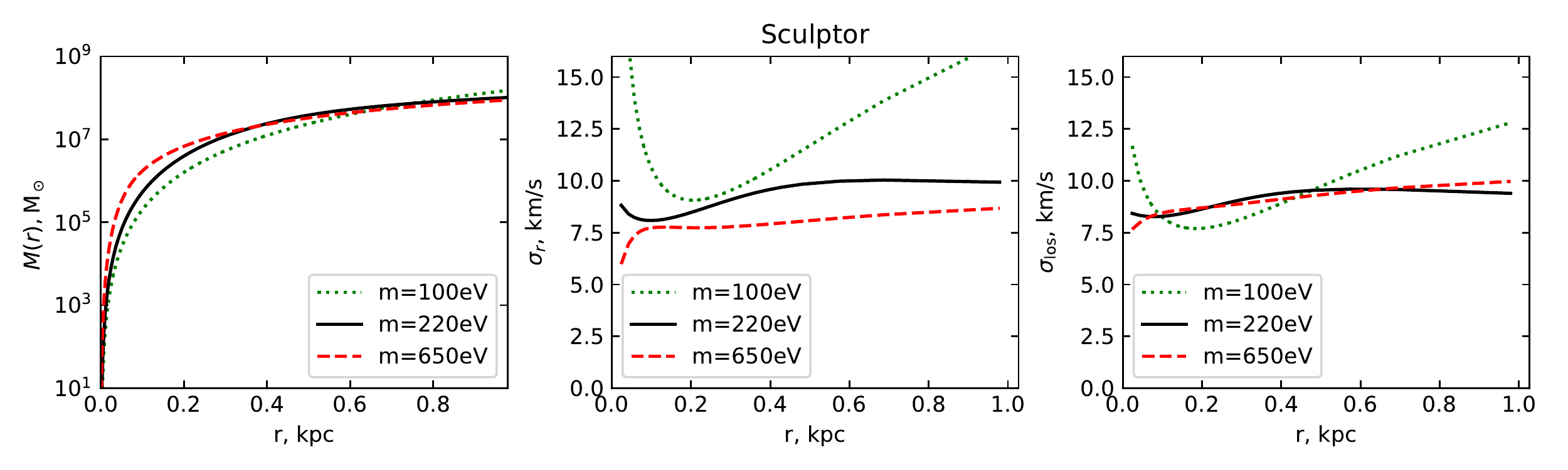}
    \includegraphics[width=\textwidth]{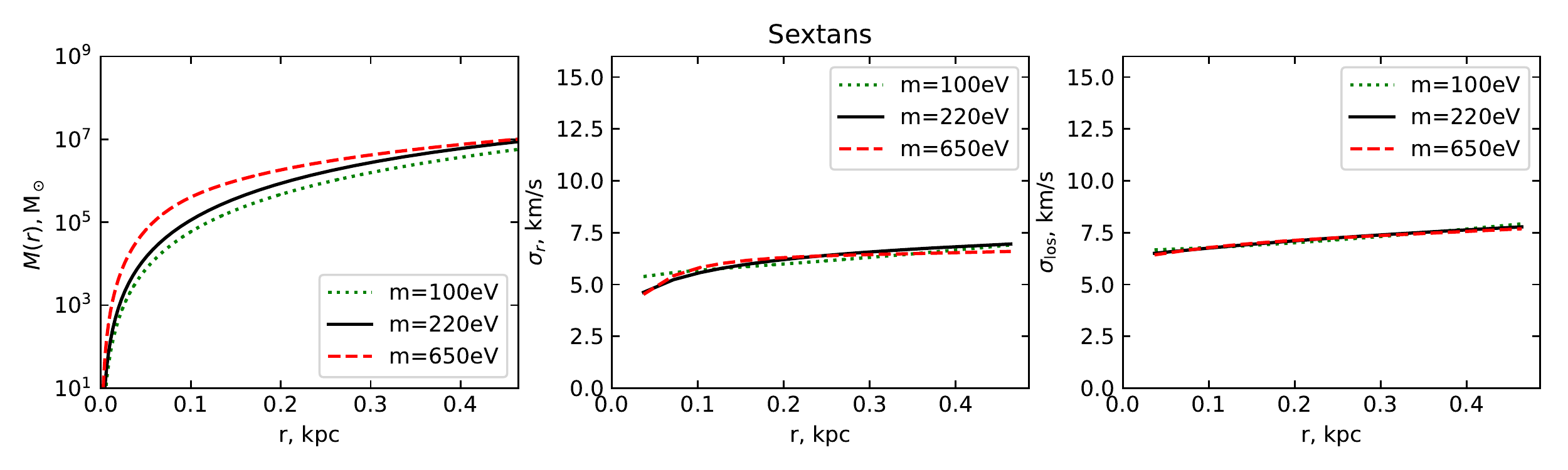}
    \includegraphics[width=\textwidth]{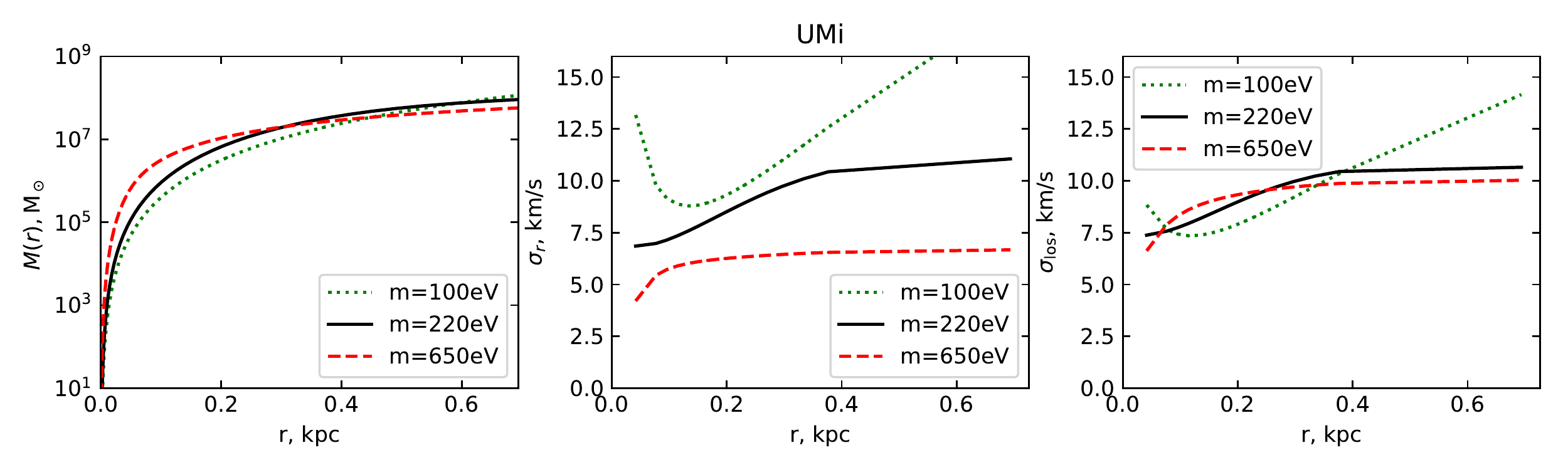}
    \caption{Same as in Fig.~\ref{fig:mass-mr-mlos-1} but for another four objects.}
    \label{fig:mass-mr-mlos-2}
\end{figure*}

The behaviour of $\sigma_\mathrm{los}$ is determined by the behaviour of  $\sigma_\mathrm{r}$, which is smoothed on the characteristic scale $r_h$ via integral transformation according to Eq.\,\ref{eq:sigmalos}, see  Fig.~\ref{fig:mass-mr-mlos-1},~\ref{fig:mass-mr-mlos-2}. Therefore, in the following discussion we will focus on the behaviour of the radial velocity dispersion.

 Eq.\,\protect\ref{eq:sigmar} could be rewritten in the form  analogous to Eq.\,14 of \cite{DiPaolo:2017geq}:
 \begin{equation}\label{eq:logder}
    \frac{\partial\mathrm{ln}\sigma_\mathrm{r}^2}{\partial \mathrm{ln}r}=-\frac{1}{\sigma_\mathrm{r}^2}\frac{G M(r)}{r}-\frac{\partial\mathrm{ln}n_\star}{\partial\mathrm{ln}r}-2\beta\, . 
 \end{equation}
 According to this equation, the logarithmic slope of $\sigma_\mathrm{r}$ depends on three different terms. The first negative term dominates on large scales. On the scales $\gtrsim r_\mathrm{h}$, the influence of the second positive term  $-\frac{\partial\mathrm{ln}n_\star}{\partial\mathrm{ln}r}=5 r^2 / (r^2+r_h^2)$ is also significant. A density profile with a few-kpc core is similar to a constant-density profile for $r\ll r_\mathrm{c}$, and such halo has much lower mass enclosed into radii  $\lesssim 1$ kpc compared to a more cusped one (see Fig.~\ref{fig:mass-mr-mlos-1},~\ref{fig:mass-mr-mlos-2}). 
 In this case the first term in Eq.~\ref{eq:logder} is larger on the scales $\lesssim 1$ kpc compared to the case of more dense halos. Therefore, the logarithmic slope of $\sigma_\mathrm{r}$ is larger for halos built from low-mass fermions on the scales $r_\mathrm{h} \lesssim r \lesssim r_\mathrm{c}$. Such slope is not compatible with the data, and could be partially corrected by the third term of the Eq.~\ref{eq:logder} with positive $\beta$. 
However, large positive $\beta$ leads to fast decreasing profile of $\sigma_\mathrm{r}$ for such halos in the low-$r$ region\footnote{The asymptotic behaviour of the radial velocity dispersion is $\sigma_\mathrm{r}\sim  r^{-2\beta} + C r^2$ in the region of small $r$ for cored halos, where $C$ is some constant.}.
The dip produced in this case is reflected in the profile of $\sigma_\text{los}$, which also limits the ability to choose very large $\beta$.
Generally speaking, the discussed behaviour of the radial and line-of-sight velocity dispersions is reflected in the decrease of the best-fitting $\beta$ values with an increase of $r_\mathrm{c}/r_\mathrm{h}$, see Table~\ref{tab:corerh}.

Despite the large spread of the neighbouring points, the observational data can be regarded as "flat", i.e. preferring $\sigma_\text{los}$ profiles without large dips or high slope. Taking into account the discussion above, one can conclude that the halos built with low-mass particles contradict such "flatness".

In this context we must mention that Sextans has the largest scatter between the nearby observational points among other dSphs. The values in the neighbouring points often have more than 1-sigma differences. Also, the value of $r_\mathrm{h}$ in Sextans is about 0.7 kpc, or 1.5 times larger than the maximal radial span ($r_\mathrm{max}$) of the available kinematic data. As we mentioned above, $\sigma_\mathrm{los}$ is smoothed against $\sigma_\mathrm{r}$ with the characteristic radius $r_\mathrm{h}$. Because the ratio of $r_\mathrm{max}/r_\mathrm{h}$ for Sextans is the smallest among the classical dSphs, the corresponding level of $sigma_\mathrm{los}$ "smoothness" is the largest. This leads to the fact that all best-fits have close values of goodness-of-fit statistics and similar shape. In contrast, objects with sufficiently large $r_\mathrm{max}/r_\mathrm{h}$ ratios (such as Sculptor, Ursa Minor, Fornax) demonstrate the largest variations among profiles with different values of the dark matter masses.

Also note that our best-fitting parameters for Fornax $m=171$\,eV, $M_{200}=9.57\cdot10^8$\,$M_{\odot}$, $C_{200}=53$ correspond to  a profile with $r_\text{c}=0.93$\,\unit{kpc}, which is in a good agreement with \citet{Amorisco:2012rd}\footnote{Despite that  \cite{Amorisco:2012rd} used the Burkert profile, the core radius $r_0$ is defined as $\rho(r_0)=\frac{\rho_0}{4}$, which is similar to our definition}. While, in general, our analysis shows no significant preference for the cored dark matter profiles over the cusped ones (obtained in the $\Lambda CDM$ model), Draco, Fornax, Leo1 and Sculptor may give a possible hint on such preference.

The main advantage of our analysis is the combined study of several objects: we \textit{simultaneously} fit the data for seven classical dSphs. While the fits of the data of individual objects show different preferred particle masses (see Fig.~\ref{fig:tnfw_byobj}) and lead to different bounds, the combined analysis ensures robustness of the results. Moreover, when modelling several object, we are able to produce stronger bound.
For example, the strongest limit of 100\,eV in~\citet{DiPaolo:2017geq} is obtained by analysing the smallest dwarfs, whereas the analysis of the classical dwarfs only leads to the mass limit of few tens of electron-volts. 

\begin{figure}
    \centering
    \includegraphics[width=\columnwidth]{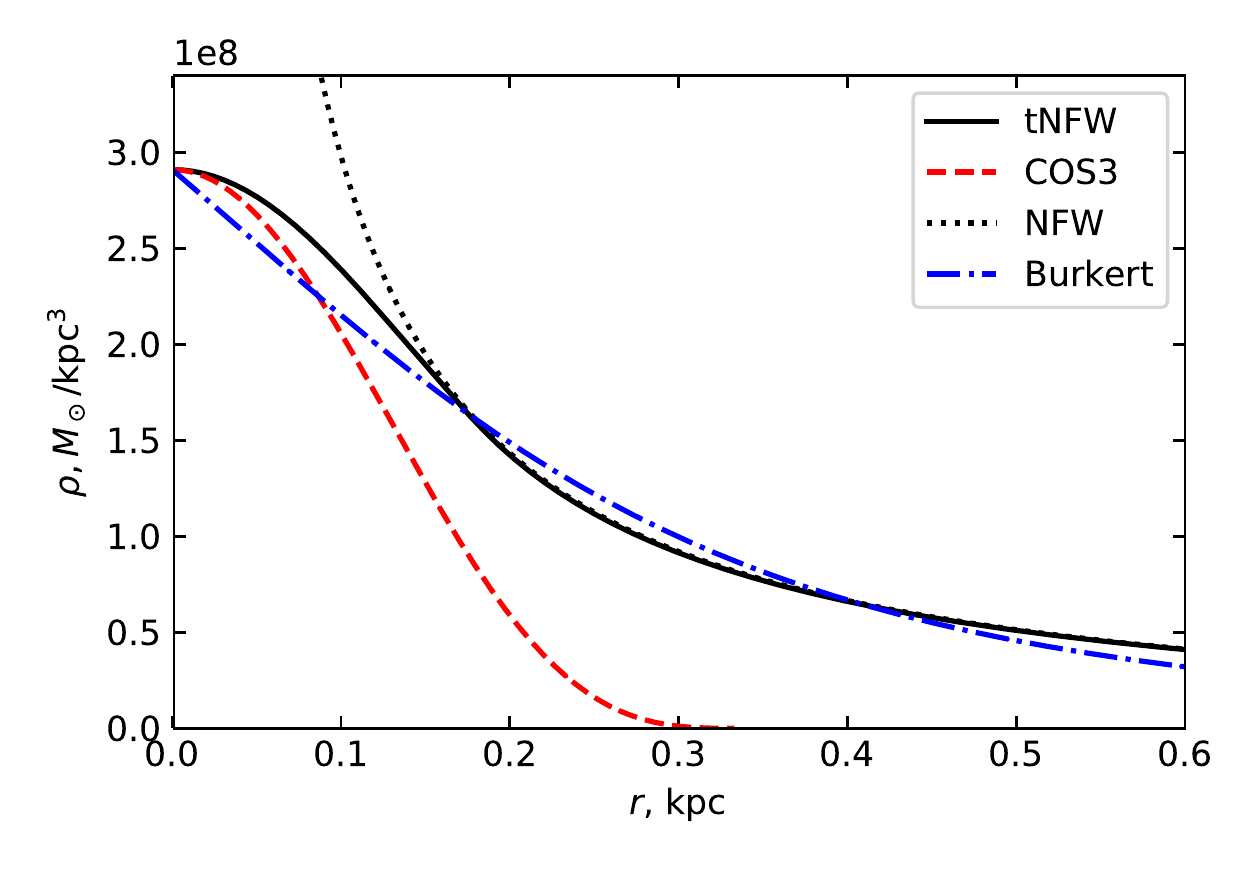}
    \caption{Dark matter density profile in the tNFW model with particle mass $m=$380\,eV, $M_{200}=1.5\times10^{10}\,M_\odot$, $c_{200}=10$ compared with the fully degenerate COS3 profile of~\protect\citet{DiPaolo:2017geq} with the same central density and particle mass. The NFW profile here has the corresponding asymptotic behaviour. Also plotted is the Burkert cored profile~\protect\citet{Burkert:95} with $r_0=0.38\,\unit{kpc}$ (twice the fully degenerate COS3 profile core size).}
    \label{fig:profiles}
\end{figure}

In general, the dark matter halo profile of~\citet{DiPaolo:2017geq} systematically prefers lower particle masses due to its fully degenerate nature, which produces sharp cut-off in the density profile. Unlike in \citet{DiPaolo:2017geq}, in our model the DM halo has two regions: a fully degenerate core and non-degenerate dispersed outskirts. Fig.~\ref{fig:profiles} shows the fast clipping of this profile and the smaller core size, compared with more ``blured'' tNFW profile. 

Recent direct measurements of 3D stellar kinematics in Sculptor \citep{Massari:2017} and kinematics data modelling via the Schwarzschild method \citep{Kowalczyk:18} revealed that the stellar velocity dispersions in the dwarf spheroidal galaxies are likely to be non-isotropic, but the uncertainties in the value of $\beta$ are very large. Therefore, we assume, for simplicity, that this quantity is constant on all radii. Inclusion of non-zero stellar velocity anisotropy into the analysis leads to a lower DM mass bound compared to the previous findings \citep[e.g.][]{Boyarsky:08a}. In the case of non-zero $\beta$, we found that DM particle masses in wide range are statistically indistinguishable. This agrees qualitatively with the results of \citet{DiPaolo:2017geq} and \citet{Randall:16} for models of non-fully degenerate fermionic halos. 
This $\beta$-degeneracy could be overcome by assuming multiple stellar sub-populations~\citep{Battaglia:2008jz,Walker:11, Agnello:2012uc,Amorisco:2012rd} or by using the Virial equations instead of the Jeans equations \citep{Richardson:2014mra}. However, the existing data, which does not include proper 3D stellar kinematics with possible asphericity of stellar populations, is not enough to completely break this degeneracy \citep{Kowalczyk:12,Genina:17,Hayashi:2018uop}.

The effects of supernova feedback  \citep{Navarro:96b,Pontzen:11,Oh:11,Teyssier:12,Zolotov:12}, other stellar feedback mechanisms~\citep{Chan:15, Onorbe:15}, and dynamical friction~\citep{ElZant:04,Sanchez:06,Romano:08,DelPopolo:15} could cause additional flattening of the dark matter profile and reduction of the central phase-space density. These mechanisms are thus degenerate with the dark-matter-induced core generation. Inclusion of these effects could increase the lower mass bound.

In the future, progress in the exploration of DM microphysics may be achieved via studying the ultra-faint dwarfs (UFDs), which are the most DM dominated galaxies that we know \citep[see, e.g.,][]{Bullock:17,Simon:19}. Their compactness also gives an opportunity to test the dark matter distribution on the smallest scales, e.g.,  dozens of parsecs. Also, the star-formation processes in UFDs should not be powerful enough to change substantially their internal density structure \citep{Onorbe:15}. However, even the most recent studies  \citep[e.g.,][]{Fritz:18, Simon:18a} allow one to obtain spectra only for only dozens of stars in the ultra-faint Milky Way satellites (unlike `classical' dwarfs, where spectra of hundreds or thousands of stars are measured). These data are not enough to obtain any detailed line-of-sight velocity dispersion profile. Lengthy observations on $\sim10$~m or planned extremely large telescopes may obtain the spectra of many more stars~\citep{Strigari:18,Weisz:2019bkv,Drlica-Wagner:19,Simon:19b}.

\begin{table}
    \centering
    \begin{tabular}{lcccccc}
Object	&	m, eV	&	$m_{200}$	&	$c_{200}$	&	$\beta$	&	$r_\mathrm{c}$, kpc	&	$r_\mathrm{h}$, kpc \\
\hline
Carina	&	100	&	35.98	&	120	&	0.74	&	2.11	&	0.25\\
Carina	&	220	&	119.4	&	6	&	0.52	&	0.99	&	0.25\\
Carina	&	650	&	111.7	&	5	&	0.19	&	0.20	&	0.25\\
\\
Draco	&	100	&	69.47	&	120	&	0.67	&	1.80	&	0.221\\
Draco	&	220	&	119.4	&	12	&	0.40	&	0.78    & 0.221\\
Draco	&	650	&	717.4	&	6	&	0.10	&	0.17    & 0.221\\
\\
Fornax	&	100	&	33.92	&	200	&	0.41	&	2.10    & 0.710\\
Fornax	&	220	&	10.93	&	26	&	-0.24	&	0.67    & 0.710\\
Fornax	&	650	&	13.34	&	19	&	-0.50	&	0.15    & 0.710\\
\\
Leo1	&	100	&	52.27	&	200	&	0.83	&	1.87    & 0.251\\
Leo1	&	220	&	145.7	&	9	&	0.58	&	0.86    & 0.251\\
Leo1	&	650	&	127.6	&	8	&	0.27	&	0.18    & 0.251\\
\\
Leo2	&	100	&	76.32	&	85	&	1.00	&	1.82    & 0.176\\
Leo2	&	220	&	111.7	&	12	&	0.85	&	0.78    & 0.176\\
Leo2	&	650	&	127.6	&	9	&	0.61	&	0.17    & 0.176\\
\\
Sculptor	&	100	&	52.27	&	200	&	0.64	&	1.87    & 0.283\\
Sculptor	&	220	&	6.010	&	59	&	0.10	&	0.59    & 0.283\\
Sculptor	&	650	&	11.68	&	19	&	-0.38	&	0.15    & 0.283\\
\\
Sextans	&	100	&	25.91	&	77	&	-0.05	&	2.34    & 0.695\\
Sextans	&	220	&	145.7	&	4	&	-0.20	&	1.11    & 0.695\\
Sextans	&	650	&	875.6	&	2	&	-0.38	&	0.22    & 0.695\\
\\
UMi	&	100	&	80.55	&	160	&	0.69	&	1.70    & 0.181\\
UMi	&	220	&	6.012	&	160	&	0.06	&	0.52    & 0.181\\
UMi	&	650	&	4.311	&	38	&	-1.64	&	0.12    & 0.181\\
 
    \end{tabular}
    \caption{Best-fitting values of the halo profile parameters and halo core size $r_\mathrm{c}$ for three fixed dark matter particle masses: 100\,eV, 220\,eV, and 650\,eV. Half-light radii $r_h$ are also given for reference. The $m_{200}$ values are in $10^8 M_\odot$.}
    \label{tab:corerh}
\end{table}

\section*{Acknowledgements}

The authors are grateful to D.~Iakubovskyi and Yu.~Shtanov for collaboration and valuable comments. We thank the anonymous Referee for the comments that significantly improved the quality of the paper. This work was supported by the grant for young scientist's research laboratories of the National Academy of Sciences of Ukraine. The work of A.R. was also partially supported by the ICTP through AF-06.

%%%%%%%%%%%%%%%%%%%%%%%%%%%%%%%%%%%%%%%%%%%%%%%%%%



\bibliographystyle{mnras}
\bibliography{refs} % if your bibtex file is called example.bib

% %%%%%%%%%%%%%%%%%%%%%%%%%%%%%%%%%%%%%%%%%%%%%%%%%%

% %%%%%%%%%%%%%%%%% APPENDICES %%%%%%%%%%%%%%%%%%%%%

% \appendix

% \section{Some extra material}

% If you want to present additional material which would interrupt the flow of the main paper,
% it can be placed in an Appendix which appears after the list of references.

% %%%%%%%%%%%%%%%%%%%%%%%%%%%%%%%%%%%%%%%%%%%%%%%%%%

% Don't change these lines
\bsp	% typesetting comment
\label{lastpage}
\end{document}